\documentclass[a4paper]{article}
\usepackage{graphicx}
\usepackage{dcolumn}
\usepackage{eucal}
\usepackage{amsfonts}
\usepackage{latexsym}
\usepackage{epsfig}
\usepackage{color}

\usepackage{amsfonts,amssymb,eucal}
\usepackage{amsthm}
 \usepackage{amsmath}
\usepackage{amscd}
 \usepackage{latexsym} 
 
\begin{document}
\title{Unitarity condition  method for   global fits of the Cabibbo-Kobayashi-Maskawa matrix}
\author{Petre Di\c t\u a$^{1}$\footnote {dita@zeus.theory.nipne.ro}\\
 Institute of Physics and Nuclear Engineering,\\
P.O. Box MG6, Bucharest, Romania}
\maketitle
\begin{abstract}\noindent
We report on an exact method for  global fits of the CKM matrix by using the necessary and sufficient conditions the data have to satisfy in order to find a unitary matrix compatible with them, and this method can be applied to both quark and lepton CKM matrices. The key  condition writes  $-1\le \cos\delta\le 1$ where $\delta$ is the phase that encodes the  CP violation, and it is obtained when one describes the CKM matrix in terms of four moduli of its entries.
We use it to reconstruct many approximate unitary matrices from the PDG data, and by using the double stochasticity property of the unitary matrices we find a strong dependence of $\sin 2\beta$ on $|U_{ub}|$, that implies also strong
 correlations between $|U_{ij}|$ themselves, proving the constraining power of this approach.

\end{abstract}
\section{Introduction}

Recently \cite{Di} we have introduced a unitarity condition method for  constraining the Cabibbo-Kobayashi-Maskawa (CKM) matrix entries that parameterize
the weak charged current interactions of quarks in the Standard Model.
 The determination of the CKM matrix  is a lively subject in particle physics and  the determination of its four independent parameters  that govern all flavor changing transitions of quarks and leptons in the Standard Model is an
important task for both experimenters and theorists. The large interest in the subject is also reflected by the workshops organized in the last  years whose main subject was the CKM matrix \cite{BBGS, AB, BFKS, SD}.

Although it seems that at the level of experts there exists a consensus concerning  the method of  extracting from experimental data information about the CP violating phase and the invariant angles of the unitarity triangles \cite{BBGS}-\cite{Bo}, the usual method of considering only  the orthogonality of the first and third columns has some weak points  as it was  shown in \cite{Di}. Here we give more  arguments in favor of our  novel approach; the main point of the unitarity method consists in   taking as independent parameters four moduli of the CKM matrix entries, and finding from them constraints on the phase $\delta$. An alternative approach could be the use of four independent invariant  phases \cite{AKL}, or any combination of moduli and phases, the sole condition to be fulfilled being their independence. In fact the standard unitarity triangle method \cite{CL}-\cite{Bo} as it is used in the present days allows  the determination of only two independent invariant phases, showing that using only one orthogonality relation is not enough for a complete determination of the CKM matrix. Moreover we have constructed in \cite{Di} a toy model matrix that satisfies exactly what we call the double stochasticity property, see below, its first and third columns are orthogonal, and the matrix  is {\em not unitary}. Until now the toy matrix was considered as an academical curiosity, but in fact it stresses
  the necessity to consider at least another unitarity triangle for the determination of others two phases, and a  good candidate   could be that  determined by the orthogonality of the first and the third rows. 

 Our approach provides 
the necessary and sufficient condition the data have to satisfy in order to find a unitary matrix compatible with them; this condition is given by
$-1\leq\cos\delta\leq 1$, where $\delta$ is the phase entering the CKM matrix.
In section 2  we construct our  theoretical model and give a few numerical examples that show the constraining power of our approach. In section 3 we make a comparison between the standard unitarity triangle approach and ours and do some Monte Carlo simulations stressing the numerical problems that have to be solved when doing global fits. In section 4 we 
 test
 the  unitarity property  of the experimental data  by using only the information provided by the Particle Data Group (PDG) \cite{Ha}-\cite{Ed}, and  we found that there are many unitary  matrices leading to a phase $\delta$ close to  $\pi/2$. We use the double stochasticity property and the unitary matrices found in the fits to construct continuous families of unitary matrices, and as an unexpected result we uncover an explicit dependence of $\sin 2\beta$ on $|U_{ub}|$. Thus we can use the experimental data on  $\sin 2\beta$, see e.g. \cite{HF}, and the mean unitary matrices obtained from Hagiwara et al. data \cite{Ha}, and respectively Edelman et al. data \cite{Ed}, to determine unitary matrices compatible with all these data, by reversing in some sense the usual way  of getting bounds for  $\sin 2\beta$ from fits by using  the moduli $|U_{ij}|$. Working in this way all the found  matrices as well as their $\pm 1\sigma$ companions  are  unitary. The paper ends by Conclusions.

\section{Theoretical model}

 We use the standard parameterization advocated by  PDG that we write it in a
little different form 
\begin{eqnarray}
U=
\left(\begin{array}{ccc}
c_{12}c_{13}&c_{13}s_{12}&s_{13}\\
-c_{23}s_{12}e^{-i \delta}-c_{12}s_{23}s_{13}&c_{12}c_{23}e^{-i \delta}-s_{12}s_{23}s_{13}&s_{23}c_{13}\\
s_{12}s_{23}e^{-i\delta}-c_{12}c_{23}s_{13}&-c_{12}s_{23}e^{-i \delta}-s_{12}c_{23}s_{13}&c_{23}c_{13}
\end{array}\right) \label{ckm}
\end{eqnarray}
with $c_{ij}=\cos \theta_{ij}$ and  $s_{ij}=\sin \theta_{ij}$
for the generation labels $ij=12, 13, 23$, and $\delta$ is the phase that encodes the breaking of the $CP$-invariance. 
The re-phase invariance is equivalent with choosing the phases of five arbitrary CKM matrix entries equal to zero, and we  used this property for rewriting the standard parameterization \cite{CK} in the above simplest form. This form depends on four phases, those of $U_{21},\,U_{22},\,U_{31}$ and $U_{32},$ instead of five as in the usual approach.

From experiments one measures a matrix whose entries are positive
\begin{eqnarray}
V=\left(\begin{array}{ccc}
\vspace*{1mm}
V_{ud}^2&V_{us}^2&V_{ub}^2\\
\vspace*{1mm}
V_{cd}^2&V_{cs}^2&V_{cb}^2\\

V_{td}^2&V_{ts}^2&V_{tb}^2\\
\end{array}\right)\label{pos}
\end{eqnarray}
entries that, in principle, can be determined from the weak decays of the relevant quarks, and/or from deep inelastic neutrino scattering \cite{Ha}-\cite{Ed}. As the notation suggests  we make a clear distinction between the unitary CKM matrix $U$ and the {\em positive} entries  matrix $V$ provided by the data. Of course the data are in general given in terms of some functions $f_k(V_{ij}),\,k=1,\dots,N$, that depend on the $V$ entries.

The main theoretical problem is to see if from a matrix as (\ref{pos}) one can reconstruct a unitary matrix as (\ref{ckm}). For that we need a relationship between the theoretical object (\ref{ckm}) and the experimental data (\ref{pos}), and it is 
\[V=|U|^2\]
that leads to the following relations
\begin{eqnarray}
V_{ud}^2&=&c^2_{12} c^2_{13},\,\, V_{us}^2=s^2_{12}c^2_{13},\,\,V_{ub}^2=s^2_{13}\nonumber \\
 V_{cb}^2&=&s^2_{23} c^2_{13},\,\,
 V_{tb}^2=c^2_{13} c^2_{23},\nonumber\\
V_{cd}^2&=&s^2_{12} c^2_{23}+s^2_{13} s^2_{23} c^2_{12}+2 s_{12}s_{13}s_{23}c_{12}c_{23}\cos\delta,\nonumber\\
V_{cs}^2&=&c^2_{12} c^2_{23}+s^2_{12} s^2_{13} s^2_{23}-2 s_{12}s_{13}s_{23}c_{12}c_{23}\cos\delta\label{unitary},\\
V_{td}^2&=&s^2_{13}c^2_{12}c^2_{23}+s^2_{12}s^2_{23}-2 s_{12}s_{13}s_{23}c_{12}c_{23}\cos\delta\nonumber,\\
V_{ts}^2&=&s^2_{12} s^2_{13} c^2_{23}+c^2_{12}s^2_{23} +2 s_{12}s_{13}s_{23}c_{12}c_{23}\cos\delta\nonumber
\end{eqnarray}

From the relations (\ref{unitary}) one gets
\begin{eqnarray}
\sum_{i=d,s,b} V_{ji}^2-1=0, \quad j=u,c,t\nonumber \\
\sum_{i=u,c,t} V_{ij}^2-1=0, \quad j=d,s,b \label{sto}
\end{eqnarray}
that is the weakest form of unitarity.
We stress that the above relations does not test the unitarity, as it is 
usually stated in many papers; they are  necessary but not sufficient conditions. The class of positive matrices satisfying  Eqs.(\ref{sto}) is considerable larger than the class of positive matrices coming from unitary matrices. The set  (\ref{sto}) is known in the mathematical literature as doubly stochastic matrices, and the subset coming from  unitary matrices $V_{ij}^2=|U_{ij}|^2$ is known as unistochastic ones \cite{MO}. The double stochastic matrices have an important property, they are a convex set, i.e. if $V_1$ and $V_2$ are doubly stochastic so  is their convex combination $\alpha\,V_1+(1-\alpha)V_2$, $\alpha \in [0,1]$, as it is easily checked.

The first problem to solve is to find a necessary and sufficient criterion for discrimination between the two sets. The above relations provides   the necessary and sufficient conditions 
the data have to satisfy in order  the matrix (\ref{pos}) comes from a unitary matrix, and the most constraining  condition is
\begin{eqnarray}
 -1\leq\cos\delta\leq 1\label{unit1}
\end{eqnarray}
Since in relations (\ref{unitary}) $\delta$ enters only in the cosine function we can take $\delta\in[0,\pi]$ without loss of generality.
Using the central numerical values from PDG data \cite{Ha}-\cite{Ed}, or from the published fits \cite{HLLL}-\cite{JC}, one finds values for 
$\cos\delta$ outside
the physical range $[-1,1]$, or even  complex as it was shown in \cite{Di}.
The last four relations (\ref{unitary}) provide us formulas for $\cos\delta$ and these formulas have to give the same number when comparing theory with experiment, by supposing the data come from a unitary matrix. Their explicit form depends on the independent four parameters  we choose to parameterize the data, and we will always choose these parameters as four experimentally measurable quantities, i.e. $V_{ij}^2$. In fact there are 57 independent groups of four independent moduli that lead to 165 different expressions for  $\cos\delta$.  Depending on the explicit choice of the four independent parameters we get one, two, three or four  different expressions for $\cos\delta$; e.g. if we take $V_{ud},\,V_{us},\,V_{cd},\,V_{cs},$ as independent parameters we get

\begin{eqnarray}
s_{12}=\frac{V_{us}}{\sqrt{V_{ud}^2+V_{us}^2}}\,,\,\, s_{13}=\sqrt{1-V_{cd}^2-V_{cs}^2},\,\,\,{\rm and},\,\,
s_{23}=\frac{\sqrt{1-V_{cd}^2-V_{cs}^2}}{\sqrt{V_{ud}^2+V_{us}^2}}
\label{mix}
\end{eqnarray}

\noindent
and from the sixth Eqs.(\ref{unitary}) we have
\begin{center}
\begin{eqnarray}
\begin{array}{l}
\\ 
\cos\delta=\end{array}
\begin{array}{l}
V_{ud}^2+V_{cd}^2 V_{ud}^2+V_{cs}^2 V_{ud}^2 +V_{ud}^4- V_{cs}^2 V_{ud}^4
 \\
\underline{+ V_{us}^2 -V_{cd}^2 V_{us}^2 - V_{cs}^2 V_{us}^2   + V_{cd}^2 V_{ud}^2 V_{us}^2 -  V_{us}^4   -
 V_{cs}^2 V_{ud}^2 V_{us}^2 +V_{cd}^2 V_{us}^4}\\
{2 V_{ud} V_{us} \sqrt{1-V_{cd}^2-V_{cs}^2}\,\sqrt{1-V_{ud}^2-V_{us}^2}\,
\sqrt{V_{ud}^2+V_{us}^2+V_{cd}^2+V_{cs}^2-1}}\end{array}\label{ud}
\end{eqnarray}
\end{center}

 In this case, other two independent formulas are given by the last two equations in relations (\ref{unitary}), and they have to give (almost) identical numerical results if the data are compatible with unitarity. In general the data will give different numerical values for the three functions expressing  $\cos\delta$. If the independent parameters are $V_{ud},V_{ub},V_{cd},V_{cb}$, i.e. we use the information contained in the first and the third columns as does the
standard unitarity triangle, we obtain four different expressions for  $\cos\delta$. In this case the mixing angles $\theta_{ij}$ are given by 

\begin{eqnarray}
s_{12}=\frac{\sqrt{1-V_{ud}^2-V_{ub}^2}}{\sqrt{1-V_{ub}^2}},\,\,
s_{13}=V_{ub},\,\, {\rm and}\,\,
s_{23}=\frac{V_{cb}}{\sqrt{1-V_{ub}^2}}
 \label{mix1}
\end{eqnarray}
and we obtain  for $\cos\delta$ the following formula
\begin{center}
\begin{eqnarray}
\cos\delta=~~~~~~~~~~~~~~~~~~~~~~~~~~~~~~~~~~~~~~~~~~~~~~~~~~~~~~~~~~~~~~~~~~~~~~~~~~~ \nonumber\\
\frac{V_{cd}^2-V_{cb}^2 V_{ub}^2-2 V_{cd}^2 V_{ub}^2+V_{cb}^2 V_{ub}^4+V_{cd}^2 V_{ub}^4  - V_{us}^2 +
V_{cb}^2 V_{us}^2+ V_{ub}^2 V_{us}^2 + V_{cb}^2 V_{ub}^2 V_{us}^2}
{\,2\, V_{us}\, V_{ub}\, V_{cb}\,\sqrt{1-V_{ub}^2-V_{us}^2}\,\,\sqrt{1-V_{ub}^2-V_{cb}^2}}\label{uny}
\end{eqnarray}
\end{center}
Looking at equations  (\ref{mix})-(\ref{uny}) we see that the expressions defining the mixing angles and phase $\delta$ are quite different. Thus if the data are compatible to the existence of a unitary matrix these  angles $s_{ij}$ and phases $\delta^{(i)}$ have to be equal, and these are  the most general necessary  conditions for unitarity; they  can be written as
\[0\le s_{ij}^{(m)}\le 1, \,\, s_{ij}^{(m)}=s_{ij}^{(n)},\,\,m, n=1,\dots,57,\,\, \cos\delta^{(i)}=\cos\delta^{(j)},\,i,j=1,\dots,165\]
The  above relations are also satisfied  by the double stochastic matrices, and the condition that separates the unitary matrices from the  double stochastic ones is given by the relation (\ref{unit1}), i.e. $-1\le \cos\delta^{(i)}\le 1$.

As a warning, what we said before can be summarized as follows: the unitarity property is a property {\em of all the CKM matrix elements} and not the property of a row and/or a column, and it implies strong correlations between $U_{ij}$.

To better understand the above considerations, let us consider the most general exact form of a double stochastic matrix that is obtained when we assume that the relations (\ref{sto}) are exactly satisfied, or in other words  the $V$ entries are known with infinite precision. Because almost all the unitarity triangle fits \cite{BBGS}-\cite{JC} use $U_{us},\,U_{ub},\,\,{\rm and}\,\, U_{cb}$  as input parameters we make the following notation
\[|U_{us}|=a,\,\,|U_{ub}|=b,\,\,{\rm and}\,\, |U_{cb}|=c\]
Since we need four moduli we choose as the fourth one, $|U_{cd}|=d$, and with they we form the matrix
\begin{eqnarray}
S=\left(
\begin{array}{ccc}
\sqrt{1-a^2-b^2}&a&b\\
d&\sqrt{1-c^2-d^2}&c\\
\sqrt{a^2+b^2-d^2}&\sqrt{-a^2+c^2+d^2}&\sqrt{1-b^2-c^2}
\end{array}\right)\label{toy}
\end{eqnarray}
The squared matrix $S^2$ is an exact doubly stochastic matrix as it is easily checked. If we make the substitution $V_{ij} \rightarrow S_{ij}$, all the 165 $\cos\delta$ expressions have the same form, namely
\begin{eqnarray}
\cos\delta=\frac{d^2(1-b^2)^2-a^2(1-b^2-c^2)-b^2c^2(1-a^2-b^2)}{2 a b c \sqrt{1-a^2-b^2} \sqrt{1-b^2-c^2}}\label{cosd}
\end{eqnarray}
From the condition 
\[\cos^2\delta\leq 1\]
we obtain the equivalent relation
\begin{eqnarray}\begin{array}{c}
a^4-2 a^4 c^2-2 a^2 b^2 c^2+a^4 c^4+2 a^2 b^2 c^4+ b^4 c^4-2 a^2 d^2+2 a^2 b^2 d^2+2 a^2 c^2 d^2 \\-2 b^2 c^2 d^2
+2 a^2 b^2 c^2 d^2+2 b^4 c^2 d^2+d^4-2 b^2 d^4+ b^4 d^4 \leq 0\end{array}
\end{eqnarray}
that describes the physically admissible region in the 4-dimensional space generated by the moduli $a,b,c,d$. For a similar result see \cite{Br}, where the authors proposed, for the first time, a parametrization of the CKM matrix in terms of four independent squared moduli, and they looked for  constraints implied by unitarity. This result proves that four independent moduli do not determine a unitary matrix, even when the moduli satisfy the relations (\ref{sto}); this happens then and only then, when the relation (\ref{unit1}) is satisfied, and the above relation separates the unitary matrices from the doubly stochastic ones in our choice of independent moduli. If we select other four moduli we find another
form of the region. To see that the above constraint is very strong, let us take  the median values from \cite{Ed} for $a, b, c$, and then  we find that $\cos\delta$
takes physical values if and only if 
\[0.223659\leq d\leq 0.223958\]
i.e. a very small interval from the allowed range $0.221\leq d\leq 0.227$.
If we use instead of the central values for the above parameters the recommended  values given in \cite{Ed} and write them as rational numbers, i.e. $a=0.22=11/50,\,b=0.00367=367/100000,\, c= 0.0413=413/10000,\,d=0.224=28/125 $, one finds
\[\cos\delta=\frac{1160896633137657437088689}{4158110000000\sqrt{94995845323482088321}}\approx 28.6448\]
If we take $a=0.224$, i.e. we modify it only with the small  quantity $4\times 10^{-3}$ we find  $\cos\delta\approx 1.28$, that shows that the fulfillment of 
 unitarity imposes very strong correlations between all the entries of the CKM matrix.

Now we define a test function that has to take into account the double
stochasticity property expressed by the 
conditions (\ref{sto}) and the fact that in general the numerical values of data are such that
  $\cos\delta$ depends on the choice of the four independent parameters and could take values outside the physical range. Our proposal is
\begin{eqnarray}
\chi^2_1=\sum_{i < j}(\cos\delta^{(i)} -\cos\delta^{(j)})^2+\sum_{j=u,c,t}\left(
\sum_{i=d,s,b}V_{ji}^2-1\right)^2\nonumber\\
+\sum_{j=d,s,b}\left(
\sum_{i=u,c,t}V_{ij}^2-1\right)^2,\,\,\,\,-1\le\cos\delta^{(i)}\le 1\label{chi} 
\end{eqnarray}
This proposal is our main contribution to the existing methods for fitting the CKM matrix entries, and it expresses the full content of unitarity. We stress again that the both conditions have to be fulfilled, $\chi^2_1$ has to take small values and $\cos\delta$ values have to be physically acceptable, since   the  relation $\chi^2_1\equiv 0$ holds true for both double stochastic and unitary matrices. The second component 
that takes into account the experimental data has the form
\begin{eqnarray}
\chi^2_2=\sum_{i=u,c}\,\,\sum_{j=d,s,b}\left(\frac{V_{ij}-\widetilde{V}_{ij}}{\sigma_{ij}}\right)^2\label{chi2} 
\end{eqnarray}
where $\widetilde{V}_{ij}$ is  the numerical matrix that describes the experimental data, and $\sigma$ is the matrix of errors associated to $\widetilde{V}_{ij}$. 
In the last sum we use only the data coming from the  first two rows because the entries of the third row  are not yet measured. The above form can be easily modified if we have other experimental information for some functions depending on $V_{ij}$. 

The above expressions
 can be used to test globally the unitarity property of the experimental data and of the published fits.

\section{Comparison of the two methods}

Unitarity triangle approach exploits other consequences of the unitarity property of the matrix (\ref{ckm}), namely the orthogonality relations of rows, and, respectively, columns of a unitary matrix. Although there are six such relations usually one considers only the orthogonality of the first and the third columns of $U$, relation that is written as
\begin{eqnarray}
U_{ud} U_{ub}^* + U_{cd} U_{cb}^* + U_{td} U_{tb}^*=0\label{ort}
\end{eqnarray}
The above relation  is  scaled by dividing it 
  through the middle term such that the length of one side is 1. 
The other sides have the lengths
\begin{eqnarray}
R_u&=&\left|\frac{U_{ud} U_{ub}^*}{U_{cd} U_{cb}^*}\right|=\frac{b\,\sqrt{1-a^2-b^2}}{c\,d}\nonumber\\
R_t&=&\left|\frac{U_{td} U_{tb}^*}{U_{cd} U_{cb}^*}\right|=\frac{\sqrt{a^2+b^2-d^2}\,\,\sqrt{1-b^2-c^2}}{c\,d}\label{tri1}
\end{eqnarray}
where we have written on the right hand side the R-values in our choice of the four independent parameters that lead to the double stochastic matrix  (\ref{toy}). If as in the preceding case we compute the expressions on the right hand side using the  values recommended in  \cite{Ed} we find
$R_u=0.387923$, and $R_t=4.53416 i$.
If  $R_u$ is in the ``normal'' range, $R_t$ gets imaginary. Please remark that both  approaches send the same signal, although in different ways: something is wrong with the recommended  values. Using the above expressions one sees that $ R_t$ gets positive only if $a^2+b^2-d^2\ge 0$. 

\begin{figure}[th]
\centerline{\psfig{file=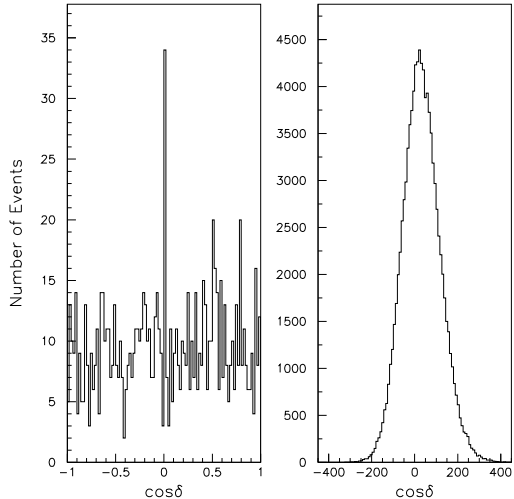,width=3.0in}}
\vspace*{8pt}
\caption{Number of Events as function of  $\cos\delta$. The input values are:
 $a=0.2200\pm0.0026,\,\,b=(3.67\pm 0.47)\times10^{-3},\,\ c=(41.3\pm 1.5)\times10^{-3},\,\, d=0.224\pm 0.012$. On the left panel one sees the number of events in the physical region,  $\cos\delta\in (1,1)$.   On the right panel are all the  events and the interval of variation is $\cos\delta\in (-350,\,\,450)$. The number of physical events is  $\approx 1\%$ of the total simulated data, and there is a peak around $\cos\delta\approx 0$.}\label{f1}
\end{figure}

\begin{figure}[th]
\centerline{\psfig{file=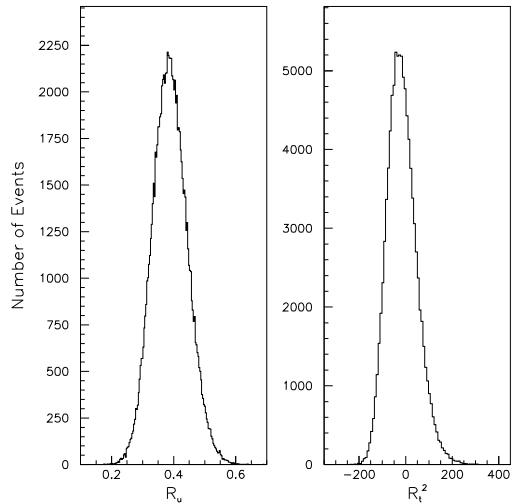,width=3.0in}}
\vspace*{8pt}
\caption{Number of Events as function of  $R_u$, and $R_t^2$. The input values are:
 $a=0.2200\pm0.0026,\,\,b=(3.67\pm 0.47)\times10^{-3},\,\ c=(41.3\pm 1.5)\times10^{-3},\,\, d=0.224\pm 0.012$. $R_u$ takes values within the interval  $R_u\in (0.2,\,\, 0.6)$, and $R_t^2\in (-250,\,\,250  )$, and only $37.5\%$ of them are positive. If we impose the triangle inequality, $|R_u -R_ t|\le 1\le R_u+R_t$,  the number of $R_t$ events is  around  $1\%$. }\label{f2}
\end{figure}

To have a quantitative estimate of the constraining power of both approaches we did a Monte Carlo simulation of the expressions appearing in (\ref{cosd}) and (\ref{tri1}) using as input $a=0.2200 \pm 0.0026,\,\,,b=(3.67\pm 0.47)\times10^{-3},\,\ c=(41.3\pm 1.5)\times10^{-3},\,\, d=0.224\pm 0.012$ and the results are shown in Fig.\ref{f1} and  Fig.\ref{f2}. The results show that only a tiny fraction of simulated data are in the physical region.

Because the data are affected by errors we have to find all the possible forms for $R_u$ and $R_t$, and for all the six unitarity triangles, and we can write a similar form for $\chi_1^2$. Thus there is a similarity between the two approaches and in {\em this form} they have to lead to similar results.

Let us see how  the problem is solved in the usual manner \cite{BLO},
 \cite{CL}, \cite{HLLL}, \cite{JC}, \cite{Bo}. Since the unitarity is {\em assumed} and {\em built} in the form (\ref{ckm}) the relation  (\ref{ort}) is an identity; this means that we have to prescribe how the experimental data are related to the $U$ entries. The first attempt was done by Wolfenstein \cite{W}, its nice feature being the possibility to estimate the order of magnitude for all $U$ entries at a time when the experimental data were scarce. Being an approximation depending on a small parameter, the unitarity is far from being exactly  satisfied. Taking into account that  there is no simple relation between the parameters defined by Wolfenstein, $\lambda,\, A,\,\rho,\,\eta$, and directly measurable quantities, Branco and Lavoura \cite{BL} proposed  that instead of $\delta$ it is better to choose another parameter,  $q=|U_{td}/(U_{us}\,U_{cb})|^2$, this one  being more relevant to $B-$mesons physics and a {\em measurable} quantity. Although the given parametrization is a Wolfenstein-type one with four independent moduli almost nobody paid attention to it.

Nowadays use of standard unitarity triangle approach is largely based on a paper by Buras et al \cite{BLO}. They tried to restore the unitarity of the CKM matrix which in Wolfenstein parametrization is not satisfied exactly. Thus they choose  four independent parameters as given by, see their Eq.(2.2),
\begin{eqnarray}
s_{12}=|U_{us}|,\quad s_{13}=|U_{ub}|,\quad  s_{23}=|U_{cb}|,\quad \delta
\label{l1}
\end{eqnarray}
and that set is replaced by 
\begin{eqnarray}
 \lambda,\quad A,\quad \rho,\quad \eta\label{l2}
\end{eqnarray}
for providing a unitary CKM matrix in terms of  Wolfenstein parameters. To this end they go back to (\ref{l1}) and impose the relations
\begin{eqnarray}
s_{12}=\lambda,\quad s_{23}=A\,\lambda^2,\quad  s_{13} e^{-i\delta} = A\,\lambda^3(\rho- i\eta)\label{l3}
\end{eqnarray}
to {\em all orders} in $\lambda$.
It follows that
\begin{eqnarray}
\rho=\frac{s_{13}}{s_{12} s_{23}} \cos\delta,\qquad \eta=\frac{s_{13}}{s_{12} s_{23}} \sin\delta,\qquad \label{l4}
\end{eqnarray}
They observe that (\ref{l3}) and  (\ref{l4}) represent simply the change of variable from (\ref{l1}) to  (\ref{l2}). Further below they say: Making this change of variables in the standard parametrization  we find the CKM matrix as a function of $(\lambda,A,\rho,\eta)$, which satisfy unitarity exactly.  So the unitarity problem of a parametrization \`a la Wolfenstein is completely solved.
However . from the relations (\ref{l4}) we get 

\begin{eqnarray}
\rho^2+\eta^2=\frac{s_{13}^2}{s_{12}^2 s_{23}^2}=\frac{V_{ub}^2}{V_{us}^2 V_{cb}^2}, \quad \tan\delta=\frac{\eta}{\rho}\label{l5}\end{eqnarray}
The above relations tell us that only the first one is expressed in terms of {\em measurable} quantities, the second one is not. Taking into account that $\delta$ is the parameter which describes directly the CP-violation it would been better to use for its determination a formula given in terms of measurable quantities. As we suggested at the beginning of this section this can be done, and a 
 detailed comparison of both the methods, as well as  its implications will be treated elsewhere \cite{Di3}. 

\section{Results}
                          
For our fits in this paper we used the same 17 independent expressions for $\cos\delta$ as in \cite{Di}, and  for improving the statistics we considered another 7 groups of four independent moduli leading to a total of 42 independent expressions for $\cos\delta$. We considered both the experimental data \cite{Ha} and \cite{Ed}, and the results have many features in common, although there are a few remarkable differences.
\begin{table}
\begin{tabular}{| c l l l l|}
\hline
Quantity &~~~ & Central value $\pm$ error&~~~& Central value $\pm$ error \\
&& Data from \cite{Ha}&& Data from \cite{Ed}\\\hline
$V_{ud}$&&0.974868 $\pm$ 0.000031&&0.974534$\pm$0.000027 \\
$V_{us}$&& 0.222717  $\pm$  0.000031&& 0.224138$\pm$0.000027 \\
$V_{ub}$&&(5.356056 $\pm$  0.035)$ 10^{-3}$&& (3.34838$\pm$0.005)$ 10^{-3}$ \\
$V_{cd}$&&0.222529 $\pm$  0.00003&&0.2240219$\pm$0.000028  \\
$V_{cs}$&&0.9740447  $\pm$  0.00005&&0.973699$\pm$0.000025 \\
$V_{cb}$& &$(4.1445  \pm 0.0064)10^{-2}$&& (4.151237$\pm 0.0013)10^{-2}$ \\
$V_{td}$ & &$(10.60047  \pm 0.033) 10^{-3}$ && $(9.862\pm 0.0042) 10^{-3}$\\
$V_{ts}$ &&$(4.042288  \pm 0.006)10^{-2} $&&$ (4.04626\pm 0.0013)10^{-2}$\\
$V_{td}$& &0.999126  $\pm$ 0.00008&& 0.999132$\pm$0.000014\\
\hline
\end{tabular}
\caption{Fit results  and errors using the standard input from PDG data \cite{Ha}-\cite{Ed}. The errors are only statistical and come from the fact that $\chi_1^2$ is not zero; for a unitary matrix the errors are zero. However they suggest the order of magnitude for the precision with which the experimental data have to be measured for a unambiguous determination of $\delta$.  The results show that there are   unitary matrices compatible with the data.}
\end{table}

Taking as test function $\chi^2=\chi^2_1+\chi^2_2$ given as in Eqs.(\ref{chi})-(\ref{chi2}) we obtained two (approximate) unitary matrices, one from the data \cite{Ha}, and the second by using the novel data  \cite{Ed}. As in \cite{Di} we relaxed the condition 
 (\ref{chi2}) by taking all combinations with five and respectively four $V_{ij}$ that provided ${6\choose 1}+{6 \choose 2}=21$ new matrices. The set of these  matrices was considered as 22 independent ``experiments'' for each experimental data set on which the statistics was done. For $\cos\delta$ these lead to
 $22\times 42= 924$ values that gave $\delta=89.979^0 \pm 0.041^0$ for Hagiwara et al data \cite{Ha}, and to  $\delta=89.9954^0 \pm 0.00023^0$ for Edelman et al data \cite{Ed}. It is remarkable that the results of the fits concerning $\delta$ are consistent with the Monte Carlo simulations, see Fig.\ref{f1}. The  $\chi^2_1$ values  are within the interval $2.\times 10^{-7}\le\chi^2_1 \le 1.7\times 10^{-3}$  and the final results concerning the moduli are shown in the  Table 1.

Looking at the final results we see that  $V_{ij}$ values are not  far from those provided by other fits, and are around the central values of both experimental data sets.
 The only unusual  feature concerns the mean values 
for $V_{ub}$, both being more or less far from the experimental values.  In fact the intervals of variation were $1.9\times 10^{-4}\le V_{ub}\le 3.72\times 10^{-3} $, for Edelman et al. data \cite{Ed}, and respectively,  $3.29\times 10^{-3}\le V_{ub}\le 11.6\times 10^{-3}$, for Hagiwara et al. data \cite{Ha}.  The essential difference between the PDG data comes from the central values of the parameters showing that these ones cause two different behaviors for functions depending on $V_{ij}$. But the most   striking feature is the explicit strong dependence of the angle $\beta$ on $|U_{ub}|$. For seeing it we  used the double stochasticity property and with the mean 
 $\sqrt{V}$ matrices $H$ and $L$ given in Table 1, obtained from Hagiwara et al. data, and, respectively, Edelman et al. data, we form the convex combination 
\begin{eqnarray}
A=\sqrt{x H^2+(1-x)L^2}\label{sot} 
\end{eqnarray}

\begin{center}
\begin{figure}
\epsfig{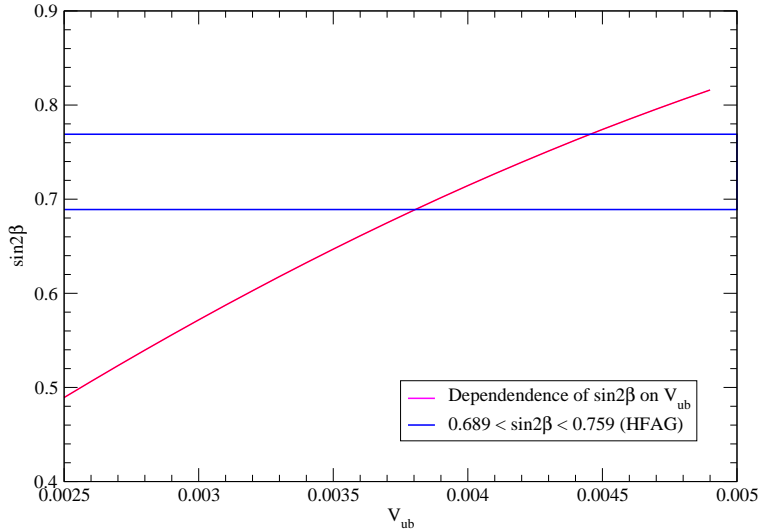}
\caption{ $\sin 2\beta$ as function of $V_{ub}$; the horizontal band
  corresponds to $\sin 2\beta =0.726\pm 0.037$ data,  as given in \cite{HF}. HFAG results imply $3.84456\times 10^{-3}\le |U_{ub}|\le 4.40217\times 10^{-3}$.}
\end{figure}\end{center}
matrix that depends on a continuous parameter. This dependence can be expressed in terms of any  $V_{ij}$, at our choice. If we choose $V_{ub}$ as free parameter we find for
 $\sin{2\beta}$ the behavior shown in Fig.3. Taking the HFAG value \cite{HF} for granted, $\sin{2\beta}=0.726\pm 0.037$,
we find  $0.00380456\le V_{ub}\le 0.00440217$ and the central value $V_{ub}=
0.00409197$. Substituting them in Eq.(\ref{sot}) we find three
 matrices, all of them being compatible to unitarity, the central one being
\begin{eqnarray}
\sqrt{V}=\left(\begin{array}{ccc}
0.97464&0.223741&0.00409197\\
0.22355&0.973809&0.0414911\\
0.0101019&0.04045&0.99913
\end{array}\right)
\end{eqnarray}
These results imply very strong correlations between all $V_{ij}$ and give small compatibility intervals for all $V_{ij}$, e.g. we get
$0.223542\le |U_{us}|\le 0.223935$, 
$0.0414821\le |U_{cb}| \le 0.0414998$, $0.040447\le |U_{ts}|\le 0.0404552$, etc. Thus our approach shows the constraining power of the unitarity method and allows us  to obtain matrices compatible to unitarity and the experimental data in a direct way.

\section{Conclusions}

What we have done in the paper  is a ``numerical experiment'' using the PDG data and the unitarity method for constraining the data, experiment that has shown a unexpected behavior of  physical quantities,  behavior that cannot be seen by using the previous methods for doing fits.  As a conclusion we may  say  that our approach to the reconstruction of CKM unitary matrices being exact and using four independent moduli can be applied to both quark and lepton matrices and could outperform by far all the other methods used to reconstruct a CKM unitary matrix from experimental data. Preliminary results on the lepton matrix $\sqrt{V}$ \cite{GG} are very interesting and will be treated elsewhere. 

In the same time our approach seems to be the ideal tool for a coherent treating  of all the experimental data available on moduli and angles of the unitarity triangles allowing to obtain results as those shown in Fig.3.

On the other hand the test $\chi^2_1\approx 0$, associated to physical values for $\cos\delta$, that expresses the full content of unitarity, has to be satisfied by all the present  and future fits, and  it have to be taken seriously into account by all the people working on unitarity triangle problems.

 All of us  have to encourage a wider appreciation of different parameterizations of unitary CKM matrices and the connections between them. Of course a selection of a specific set of angles and/or phases has no intrinsic 
theoretical significance because all the choices are {\em mathematically} equivalent; however a clever choice may shed more light on important qualitative or experimental issues.
In this sense we consider that we have to pay a special attention to the Aleksan et al. approach \cite{AKL}, whose main point was: {\em any phase-convention-independent product of CKM elements is a linear combination of such four independent phases, with  integer coefficients}, point which passed almost unnoticed. That means that all the fits have to provide values for  these  four {\bf fundamental frequencies} because the used form of the CKM unitary matrix may depend on the physical process we want to measure, by choosing it  such that the dependence of  the decay amplitude on  the CKM parameters  should be as simple as possible. Having these fundamental frequencies it will be a simple way to translate the results obtained in a specific parameterization to results obtained in another parameterization used by somebody else.
For an other phase convention different from the usual one see for example Botella et al. \cite{BB}.

 We think that  an important task for both experimenters and theorists is to
 improve all our {\em experimental and theoretical} tools we are working with, having in view the second-generation B-decay experiments at the LHC machine, and we consider our approach as a small step in this direction.

\vskip3mm
{\bf Acknowledgments}. I would like to thank the anonymous referee to the first version of the paper, who by raising a deep problem concerning a comparison between the two approaches lead me to add the new section 3. I thank also Robert Fleischer for an interesting discussion on some preliminary results presented in the paper.


\begin{thebibliography}{99}

\bibitem{Di} P Di\c t\u a, hep-ph/0408013
\bibitem{BBGS}M Battaglia, A J Buras, P Gambino, and A Stocchi (Eds.)  Proceedings of the Workshop {\em The CKM Matrix and the Unitarity Triangle},
13-16 February (2002), CERN, Geneva, hep-ph/0304132

\bibitem{AB} H Abele and D Mund (Eds.), Proceedings of the Two-Day-Workshop {\em Quark-Mixing, CKM-Unitarity}, September 19-20 (2002), Heidelberg,
hep-ph/0312124

\bibitem{BFKS} P Ball, J M Flynn, P Kluit, and A Stocchi (Eds.), Proceedings of the {\em 2nd Workshop on the CKM Unitarity Triangle}, IPPP Durham, April 2003
\bibitem{SD} CKM2005 Workshop, http://ckm2005.ucsd.edu/
\bibitem{BLO} A J Buras, M E Lautenbacher and G Ostermaier, Phys.Rev {\bf D50} (1994) 3433
\bibitem{CL} M Ciuchini et al. JHEP {\bf 0701} (2001) 013

\bibitem{HLLL} A H\"ocker, H Lacker, S Laplace, and F R Le Diberder  Eur.Phys.J. {\bf C21} (2001) 225

\bibitem{JC} J Charles  {\em et al.} (The CKM Fitter Group), hep-ph/0406184
\bibitem{Bo} M Bona et al., hep-ph/0501199
\bibitem{AKL} R Aleksan, B Kayser and D London,  Phys.Rev.Lett. {\bf 73} (1994) 
\bibitem{Ha} K Hagiwara {\em et al.} Phys.Rev. {\bf D 66} (2002) 010001
\bibitem{Ed} S Eidelman   {\em et al.} Phys.Lett. {B \bf 592} (2004) 1
\bibitem{HF} http://www.stanford.edu/xorg/hfag/triangle/
\bibitem{CK} L L Chau and W Y Keung, Phys.Rev.Lett. {\bf 53} (1984) 1802

\bibitem{MO} A W Marshall and I Olkin, {\em Inequalities: Theory of Majorization and Its Applications}, (Academic Press, New York, 1979), Chapter 2
\bibitem{Br} G C Branco and L Lavoura, Phys.Lett. B {\bf 208} 123 (1988). I thank L Lavoura for pointing out this paper to me.

\bibitem{W} L Wolfenstein, Phys.Rev.Lett. {\bf 51} 1945 (1983) 
\bibitem{BL} G C  Branco and L Lavoura, Phys.Rev. {\bf D 38} 2295 (1988)
\bibitem{Di3} P Di\c t\u a, in preparation
\bibitem{GG} M C Gonzalez-Garcia, hep-ph/0410030

\bibitem{BB} F J Bottela, G C Branco, M Nebot, and M N Rebelo, Nucl.Phys. {\bf B 651} (2003) 174
\end{thebibliography}
\end{document}